\documentclass{article}

\usepackage{arxiv}

\usepackage[utf8]{inputenc} 
\usepackage[T1]{fontenc}    
\usepackage{hyperref}       
\usepackage{url}            
\usepackage{booktabs}       
\usepackage{amsfonts}       
\usepackage{nicefrac}       
\usepackage{microtype}      
\usepackage{lipsum}		
\usepackage{graphicx}
\usepackage{natbib}
\usepackage{doi}
\usepackage{algorithm}
\usepackage{algorithmicx}
\usepackage{algpseudocode}
\usepackage{amsmath}
\usepackage{subfigure}
\usepackage{multirow}

\usepackage{color}

\title{Chunk Content is not Enough: Chunk-Context Aware Resemblance Detection for Deduplication Delta Compression}

\author{ {Xuming Ye${^\dagger}$, Xiaoye Xue${^\dagger}$, Wenlong Tian}\thanks{Coresponding Author} \\
		\thanks{Both authors contributed equally to this research}
	Department of Computer Science\\
    University of South China\\
	Hunan, China 421001 \\
	\texttt{xumingye@stu.usc.edu.cn} \\
	\texttt{20189350201@stu.usc.edu.cn} \\
	\texttt{wenlongtian@usc.edu.cn} \\

	\And
    {Zhiyong Xu} \\
	Department of Computer Science\\
    Suffork University\\
	Boston, USA 02108 \\
	\texttt{zxu@suffolk.edu} \\
	\And
    {Weijun Xiao} \\
	Department of Electrical and Computer Engineering\\
    Virginia Commonwealth University\\
	Richmond, USA 23229 \\
	\texttt{wxiao@vcu.edu} \\
	\And
    {Ruixuan Li} \\
	Department of Computer Science\\
    Huazhong University of Science and Technology\\
	Wuhan, China 430074 \\
	\texttt{rxli@hust.edu.cn} \\
}

\renewcommand{\shorttitle}{\textit{arXiv} Template}

\hypersetup{
pdftitle={A template for the arxiv style},
pdfsubject={q-bio.NC, q-bio.QM},
pdfauthor={David S.~Hippocampus, Elias D.~Striatum},
pdfkeywords={First keyword, Second keyword, More},
}

\begin{document}
\maketitle

\begin{abstract}
With the growing popularity of cloud storage, removing duplicated data across users is getting more critical for service providers to reduce costs. Recently, Data resemblance detection is a novel technology to detect redundancy among similarity. It extracts feature from each chunk content and treat chunks with high similarity as candidates for removing redundancy. However, popular resemblance methods such as "N-transform" and "Finesse" use only the chunk data for feature extraction. A minor modification on the data chunk could seriously deteriorate its capability for resemblance detection. In this paper, we proposes a novel chunk-context aware resemblance detection algorithm, called CARD, to mitigate this issue. CARD introduces a BP-Neural network-based chunk-context aware model, and uses N-sub-chunk shingles-based initial feature extraction strategy. It effectively integrates each data chunk content's internal structure with the context information for feature extraction, the impact of small changes in data chunks is significantly reduced. To evaluate its performance, we implement a CARD prototype and conduct extensive experiments using real-world data sets. The results show that CARD can detect up to 75.03\% more redundant data and accelerate the resemblance detection operations by 5.6 to 17.8 times faster compared with the state-of-the-art resemblance detection approaches.
\end{abstract}

\keywords{Resemblance Detection \and neural networks \and Deduplication \and Cloud Storage}

\section{Introduction}
With the development of network and storage technologies, cloud storage has been widely used in daily life, such as Google Drive, Dropbox, and the Baidu Cloud \cite{1}. Furthermore, people prefer to pay for their online data through the cloud storage service because of its reliability and flexibility. However, there are various redundancies among cross users' outsourced data, especially for the cloud storage scenario. These duplicate data seriously deteriorates the storage utility and increases the user's financial budget for cloud storage services. Thus, removing the duplicate data among these large volumes improve the cloud storage utilization and saves user money.

Therefore, deduplication techniques were proposed by \cite{2,3,DBLP:conf/hpcc/TianLXX18}.  As the core part of deduplication techniques, the chunking algorithm \cite{DBLP:conf/sosp/MuthitacharoenCM01} split the original data into small chunks. Then, the duplicate chunk is detected based on the hash comparison. To achieve a high deduplication ratio, the chunking algorithm is improved from the fixed-size to the content-defined chunking algorithm such as the BSW CDC \cite{DBLP:conf/sosp/MuthitacharoenCM01}, TTTD \cite{eshghi2005framework}, Elastic chunking \cite{DBLP:conf/ipccc/TianLXX17} , and Fast CDC  \cite{DBLP:conf/usenix/XiaZJFHHLZ16}. Once a rolling hash value of a data slice divided by a predefined divisor value equals zero, the data slice's end position is a chunk boundary candidate. Only the candidate close to the maximum chunk size can be selected as a chunk boundary. If there is no candidate, the chunk size is set as the maximum chunk size. However, the chunk-level hash comparison in traditional chunking algorithms hardly detects the redundancy among resemblance chunks, which contains lots of duplicate data.

To detect the duplicate data among resemblance chunks, researchers have proposed lots of resemblance detection approaches \cite{DBLP:conf/systor/AronovichABBHK09,DBLP:conf/usenix/DouglisI03,DBLP:conf/kdd/FormanEC05,DBLP:conf/nsdi/PuchaAK07,DBLP:conf/sigmod/XuPSG17}. Once two chunks are similar, only the diff part is stored by utilizing the delta compression \cite{DBLP:conf/usenix/KulkarniDLT04}. As for a popular resemblance detection method, N-transform \cite{DBLP:conf/hotstorage/ShilaneWHH12} extracts all the Rabin fingerprints \cite{rabin1981fingerprinting} of a chunk. All the Rabin fingerprints values are linearly transformed N times into N-dimensional hash sets. The top-N values, one from each of the N dimensions, are selected as features. But the feature extraction positions in N-transform have probabilities located at the tail of the chunk or the chunk's head, which significantly deteriorates the accuracy of resemblance detection. Furthermore, the N-transform suffers from the feature calculations caused by the linear transformation process.

To overcome the above problem, Finesse \cite{DBLP:conf/fast/ZhangX000W19}, the state-of-the-art resemblance detection work, calculate the chunk features with a grouping strategy. Specifically, it divides the chunk into sub-chunks and extracts their corresponding Rabin fingerprints into several contiguous sets of the same size. Then, these values construct the feature by grouping together based on each set's rank. The goal of Finesse is to achieve better performance than the N-transform in resemblance detection. But, the feature extraction scheme in Finesse only depends on the content itself and is easily impacted by modifications, which is detailed in Section\ref{Problem_Statement}. Thus, Finesse's resemblance detection accuracy is still lower than the N-transform's accuracy.

In summary, the exploration of resemblance detection is still at an elementary stage. In other words, the existing feature extraction mechanisms in the resemblance detection depends on the hash values from the content itself.  Various modification patterns in the content change these hash values and deteriorate the feature extraction. Then, the resemblance chunks are dissimilar based on their features, which is detailed in Section \ref{Problem_Statement}. Fortunately, we discover that some chunks are co-occurred in the deduplication system. If two chunk's surrounding chunks are detected as resemblance chunks, these two chunks may also be similar with high probability. We call the contiguous sequence of n chunks from a given chunk as the chunk-context. Based on this observation, combining the co-occurred surrounding chunks information with the chunk's internal structure can significantly improve the chunk representations and the resemblance detection efficiency in deduplication.


Based on our observations, we first combine the chunk-context information with the chunk content to improve resemblance detection effectiveness and efficiency. Inspired by the traditional natural language processing, we attempt to embed the chunk-context information into the feature extraction processing and propose a chunk-context aware resemblance detection scheme, called CARD. It calculates the chunk features in two steps, initial feature extraction and chunk-context embedding. By introducing the N-sub-chunk shingles, CARD generates the initial feature, which reflects the chunk content internal structure. Then, a BP-Neural network-based chunk-context aware model further embeds the chunk-context information based on these initial features. By conducting extensive experiments, CARD achieves higher accuracy and faster performance than the state-of-the-art resemblance detection schemes. The main contributions are summarized as follows:

\begin{itemize}
	\item First, we analyze the state-of-the-art resemblance detection work and present their limitations. The existing feature extraction methods in resemblance detection suffer from various modification patterns, which deteriorates similar chunk detection accuracy. Moreover, traditional resemblance detection schemes ignore the chunk-context, which plays a role in improves the effectiveness and efficiency of resemblance detection.

	\item Second, we design a chunk-context aware resemblance detection scheme, CARD, for deduplication delta compression to embed the chunk-context and the content itself into the feature. By leveraging N-sub-chunk shingles based initial feature extraction scheme and the BP-neural network based chunk-context aware model, CARD achieves a qualified chunk representation, which ensures a high resemblance detection accuracy under various modification patterns.
	
	\item Contribution 3: Finally, we conduct extensive simulations to evaluate CARD. The experimental results show that our method outperforms the state-of-the-art resemblance detection schemes. It can remove up to 75.03\% more redundancy data and accelerate the resemblance detection by 5.6$\times\sim$ 17.8$\times$ compared with the state-of-the-art resemblance detection work, N-transform and Finesse.
	
\end{itemize}

The rest of the paper is organized as follows. The related work about resemblance detection in deduplication is summarized in Section \ref{related_work}. In Section \ref{Problem_Statement}, we analyze the problems and limitations of the latest resemblance detection work in deduplication. Then, we propose a solution for resemblance detection by embedding the chunk-context information in Section \ref{Design}. Finally, we present experimental results in Section \ref{experimental_evaluation}.  In Section \ref{conclusion}, we conclude the paper.

\section{Related Work}
\label{related_work}
With the prevalence of cloud and networks, cloud storage has been widely used. There is a lot of redundancy among different users. Thus, data deduplication becomes a critical technology in data-intensive storage scenarios for eliminating these duplicate data. Obviously, less redundancy among users efficiently saves money and improves cloud storage usage. Many researchers are dedicated to the deduplication in the cloud. To simplify the description, we elaborate these related work into two categories based on whether they support removing the redundancy among the similar chunks.

\subsection{Traditional Deduplication}
Traditional deduplication remove the duplicate copies by comparing hash value for each chunk \cite{DBLP:conf/usenix/ClementsAVL09}. Many researchers focus on eliminating the duplicate data in primary and backup-level \cite{DBLP:conf/fast/LillibridgeEB13,DBLP:conf/fast/LillibridgeEBDTC09,DBLP:conf/systor/LuCGC12,DBLP:conf/usenix/FuFHHCXHL14}. Other systems use different approaches to improve the effectiveness of data deduplication, such as integrating block-level deduplication with compression \cite{2012Deduplication} and gloabal deduplication method \cite{2018Design}. Moreover, data deduplication has also attracted considerable attention for virtual machine images \cite{DBLP:conf/systor/JinM09,DBLP:conf/fast/SrinivasanBGV12,DBLP:conf/usenix/ZhaoAACTSRAB20}. However, there are lots of redundancy among non-duplicate but highly similar chunks, which can not be eliminated by traditional deduplication.

\subsection{Resemblance Detection in Deduplication}
Therefore, some researchers utilize the delta compression, a data reduction technique, to maximize the compression ratio \cite{DBLP:conf/fast/ShilaneHWH12,DBLP:conf/hotstorage/ShilaneWHH12}. It takes the delta algorithm with delta format to record the difference between similar chunks. Only the differences are recorded in delta files. Nevertheless, the delta compression induces extra computation, and I/O overheads \cite{DBLP:conf/fast/ZhuLP08}. Moreover, delta compression can hardly answer which chunks should be treated as the candidates for delta compression. 

Therefore, resemblance detection is proposed to detect the similar chunks. Aronovich el at. \cite{DBLP:conf/systor/AronovichABBHK09} propose a novel type of similarity signatures serving in the deduplication system. It combines similarity matching schemes with byte by byte comparison or hash based identity schemes. Xu el at.  \cite{DBLP:conf/sigmod/XuPSG17} propose a similarity-based deduplication system for database management systems by using byte-level encoding to achieve greater savings. There are also some other coarse-grained resemblance detection approaches \cite{DBLP:conf/usenix/DouglisI03,DBLP:conf/kdd/FormanEC05,DBLP:conf/hotstorage/ShilaneWHH12}. These methods extract features from non-overlapped chunks and may suffer from high false positives. 

To overcome the above problem,  the state-of-the-art work, such as N-transform \cite{DBLP:conf/hotstorage/ShilaneWHH12} and Finesse \cite{DBLP:conf/fast/ZhangX000W19}, attempts to improve the resemblance detection accuracy and performance using the grouping features mechanism. Both of these methods are chunk-level resemblance detection. In N-transform, it extracts the top k largest Rabin fingerprint values of a chunk. Then a super-feature of this chunk can be calculated by several such features. And Finesse extracts the largest Rabin fingerprint value in each subchunk and groups the Rabin fingerprint values as the features based on these values' ranking. However, we observe that two similar chunks also have the similar chunk-context and vice versa. The existing resemblance detection designs only consider the chunk content itself while ignoring the chunk-context's positive effect.

\section{Limitations of Previous Solutions}
\label{Problem_Statement}
\begin{figure}[htbp]
	\centering
	\includegraphics[height=4.2cm,width=8.3cm]{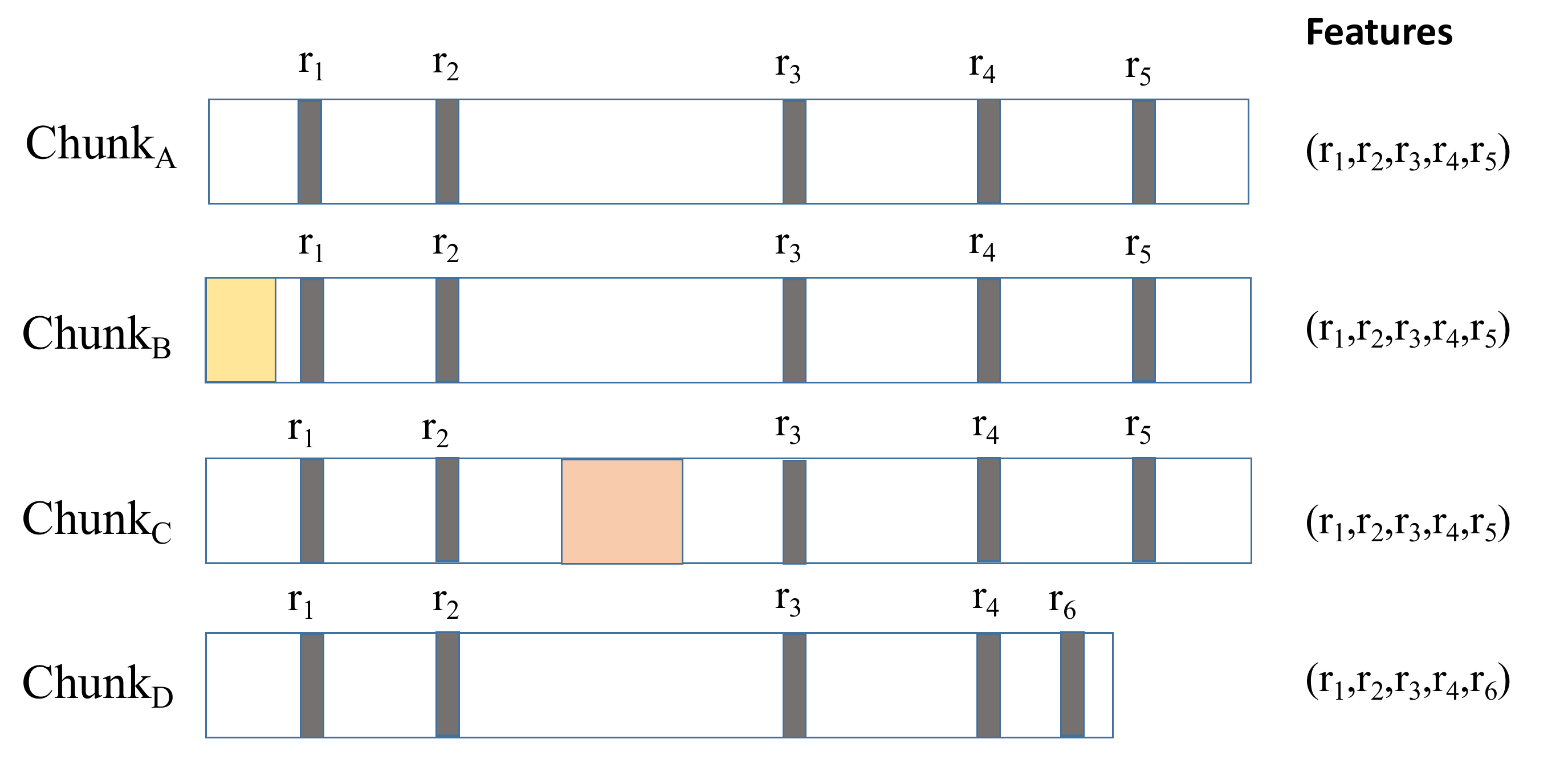}
	\caption{The Naive Chunk Feature Extraction Method}
	\label{figure_pb1}
\end{figure}

In this section, we discuss the limitations in existing resemblance detection schemes. Most researchers extract each chunk's features to detect similar chunks, as shown in Figure \ref{figure_pb1}. The common feature extraction way is to select the top k largest Rabin fingerprint values in a chunk as the feature. Then, the similarity distance could be measured by the resemblance detection. Only the different parts are stored according to the most similar chunks.

However, this naive resemblance detection method suffers a lot from modifications.  To simplify the description, we take an example to show its disadvantages. As shown in Figure \ref{figure_pb1}, we assume that $Chunk_A$ is an original version chunk. The $Chunk_B$, $Chunk_C$, and $Chunk_D$ are modification version based on $Chunk_A$ while the modification in $Chunk_B$ and $Chunk_C$ do not impact the top 5 largest Rabin fingerprint values, ($(r_1,r_2,r_3,r_4,r_5)$). The modification in $Chunk_D$ is to delete the last piece of data compared with $Chunk_A$, which contains one of the top 5 largest Rabin fingerprint value, $r_5$. Thus, the features of $Chunk_D$ is assumed as $(r_1,r_2,r_3,r_4,r_6)$. Based on this naive feature extraction,  it is hard to determine whether the $Chunk_B$ is more similar to $Chunk_A$ or $Chunk_C$ based on the features. And it also does not know whether the $Chunk_D$ is similar to $Chunk_A$ or $Chunk_B$ since the randomized character in hash value may impact the similarity distance among chunks. The situation is worse when the modification impacts the top k largest Rabin fingerprint values.


To make the resemblance detection more robust, the state-of-the-art work, such as N-transform \cite{DBLP:conf/hotstorage/ShilaneWHH12} and Finesse \cite{DBLP:conf/fast/ZhangX000W19}, attempts to improve the resemblance detection accuracy and performance using the grouping features mechanism. Specifically, it extracts the top k largest Rabin fingerprint values of a chunk. Then a super-feature of this chunk can be calculated by several such features. The differences between N-transform and Finesse are that the Finesse extracts the largest Rabin fingerprint value in each subchunk and groups the Rabin fingerprint values as the features based on these values' ranking.

\begin{figure}[htbp]
	\centering
	\includegraphics[height=4.5cm,width=8.8cm]{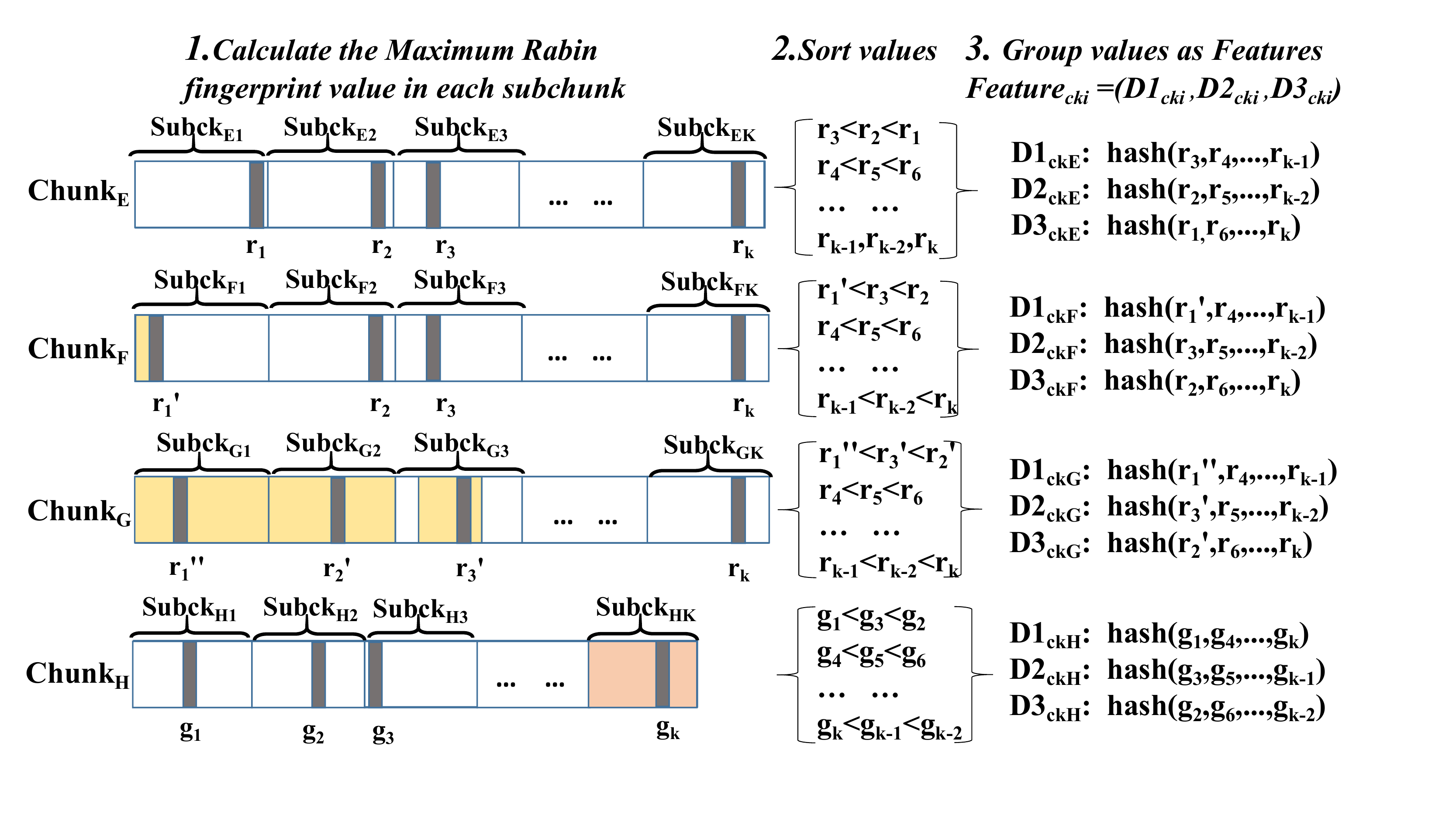}
	\caption{The Problems in Finesse under Various Modifications}
	\label{figure_pb2}
\end{figure}

As shown in Figure \ref{figure_pb2}, the Finesse divides a chunk into several sub-chunks. Then, it calculates the maximum Rabin fingerprint value in each sub-chunk. After sorting these values and dividing them into several parts, the largest Rabin fingerprint values from each part are grouped as the first feature dimension value of the chunk. The second biggest Rabin fingerprint values from each part are grouped as the second feature dimension value of the chunk, and so forth.  In Finesse, any two chunks having a feature value in common are considered highly similar and the first matched chunk is selected as the base for delta encoding \cite{DBLP:conf/usenix/KulkarniDLT04}, which is known as "FirstFit."

However, the most serious limitation in existing resemblance detection is that it does not work well under different chunk sizes. Various chunk size will result in chunks that are supposed to be similar to be considered dissimilar in high probability. What's worse, several small modifications may also result in the similar problem like naive feature extraction. To simplify the description, we take the Finesse as an example to show the above limitations in resemblance detection. As shown in Figure \ref{figure_pb2}, $subck_{Xi}$ is short for the i-th sub-chunk of $Chunk_X$. Each chunk feature has three dimensions.  $Chunk_E$ is divided into k sub-chunks where k is a predefined fixed value. Each sub-chunk has a maximum Rabin fingerprint value. Then, these values, from $r_1$ to $r_k$, are divided into sets, three values in each group, and sorted according to their values.  Finally, the feature of $Chunk_E$  is ($D1_{ckE}$,$D2_{ckE}$,$D3_{ckE}$) where $D1_{ckE}$ equals to $hash(r_3,r_4,\cdots,r_{k-1})$, $D2_{ckE}$ equals to $hash(r_2,r_5,\cdots,r_{k-2})$, $D3_{ckE}$ equals to $hash(r_1,r_6,\cdots,r_k)$.

Moreover, $Chunk_F$, $Chunk_G$ are the similar to $Chunk_E$, which modify some head information based on $Chunk_E$. $Chunk_H$ deletes some ending part based on $Chunk_E$. Based on Finesse, we can also get the features for each chunk, which is shown in Figure \ref{figure_pb2}. Even though the modification affects the maximum Rabin fingerprint value in the first sub-chunks in $Chunk_F$, the probability that $Chunk_F$'s features are different from the $Chunk_E$ features is 33.3\%. The reason is that the probability of r1 in $Chunk_E$ appearing randomly in any of the three positions is 33.3\% based on the principle of permutation and combination \cite{wiki:xxx}. The situation is worse once two similar chunks have different sizes, such as the $Chunk_H$ and $Chunk_E$.  The sub-chunk size in $Chunk_H$ and $Chunk_E$ is different under the same chunk feature dimensions. Most of the features in $Chunk_G$ are different from the $Chunk_E$ features, which are not treated as similar chunks.

In summary, the root cause of traditional resemblance detection limitations is that chunk feature only consider the chunk content itself while ignoring chunk-context. It is easily affected by various modification patterns and does not work well under different chunk sizes. However, we observe that two similar chunks may also have similar chunk-context and vice versa. These chunks are always co-occurred in the deduplication system. The chunk-context information can help the resemblance detection accuracy and decrease the modification impacts compared with traditional methods. Thus, we focus on how to combine the chunk-context information with the chunk content to achieve high-accuracy and high-performance resemblance detection in the cloud storage for deduplication.

\section{CARD Design }
\label{Design}
In this section, we first describe the overview of our design. To consider the chunk content's internal structure in the chunk representation, we propose an n-sub-chunk shingles-based initial feature extraction scheme. A BP-neural network-based chunk-context aware model is responsible for embedding the chunk-context information into the chunk initial feature to improve resemblance detection accuracy in the deduplication scenario.

\begin{figure*}[htbp]
	\centering
	\includegraphics[height=7.5cm,width=18cm]{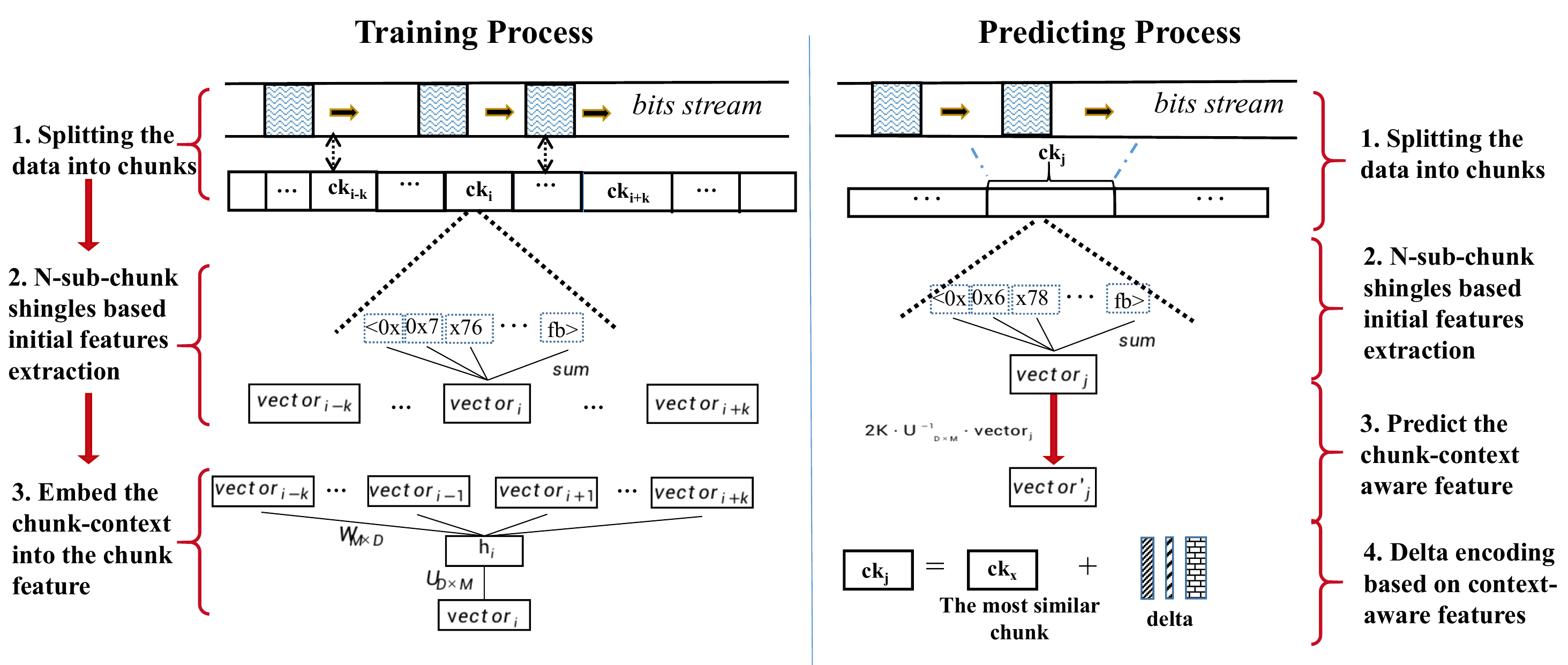}
	\caption{The CARD Workflow}
	\label{fig_scheme_workflow}
\end{figure*}

\subsection{Overview}
\label{overview}
Our primary goal is to embed the chunk-context information into the chunk representation and support the resemblance detection with the unequal chunk size. We introduce a Back-Propagation neural (BP-neural) network to reach this goal, embedding the chunk-context information into the target chunk features to achieve this goal. Compared with traditional feature extraction in resemblance detection, the chunk-context makes the chunk feature more robust under various chunk-length and modification pattern conditions. As shown in Figure \ref{fig_scheme_workflow}, we introduce our design workflow in two processes: training and predicting.


In the training process, our method first splits each file into chunks based on a traditional content defined chunking (CDC) algorithm\cite{DBLP:conf/sosp/MuthitacharoenCM01,eshghi2005framework,DBLP:conf/ipccc/TianLXX17,DBLP:conf/usenix/XiaZJFHHLZ16}. These chunks are distinct size. Next, we divide each chunk into N sub-chunks of the same size. Then, the n-sub-chunk shingles based initial feature extraction scheme, which is detailed in Subsection\ref{initial_feature_extraction_scheme}, generates an initial feature for the chunk from these N sub-chunks. Each feature has M dimensions. And the initial feature is only related to the chunk's content. We propose a BP-neural network-based chunk-context aware model to embed the chunk-context into the initial feature further. The model is generated based on the training samples, including the pair of the target chunk and its surrounding k chunk's initial features. Finally, the model continually adjusts each neural connection's weights based on these training samples.


For example, as shown in the left part of Figure \ref{fig_scheme_workflow}, the data bits stream in deduplication is split into variance chunks by traditional CDC algorithm, such as FastCDC \cite{DBLP:conf/usenix/XiaZJFHHLZ16}. We denote the i-th chunk as the target chunk, $ck_i$. Then, N-sub-chunk shingles based initial feature extraction scheme calculates each $ck_i$'s initial features, which is denoted as the $vector_i$. Then, the surrounding k chunks of $ck_i$ can be treated as the chunk-context of $ck_i$. Similarly, we also get these surrounding chunks' initial features, from $vector_{i-k}$ to the $vector_{i+k}$. Next, the $(vectors\quad of \quad the\quad  ck_i's\quad  surrounding\quad  chunks$, $vector_i)$ pairs are treated as the training data.  The surrounding chunk vectors are the chunk-context aware model's input, which is based on a kind of BP-neural network, while the target chunk vectors are treated as labels. Next,  we can get the model that transfers the $vector_i$ into the chunk-context-aware feature, $vector^{'}_{i}$ after the model training cycle iteration.  The detail of BP-neural network based chunk-context aware model is described in Subsection \ref{chunk_based context-aware model}.

The predicting process is shown in the right part of Figure \ref{fig_scheme_workflow}. We also divide the coming bits stream into chunks. And each  $ck_j$'s initial feature is denoted as $vector_j$. Then, the chunk-context aware feature $vector^{'}_{j}$ is achieved from the $vector_j$ based on the trained model. Specifically, the $vector^{'}_{j}$ equals to the $2K\cdot U^{-1}_{D\times M}$ where the matrix $U$ is a weight matrix in the trained model. Furthermore, the cloud could find the most similar chunk based on these chunk-context aware features. Next, utilizing the delta encoding technology \cite{DBLP:conf/usenix/KulkarniDLT04} removes the redundancy between these two similar chunks. Only the diff part is stored. It is noted that the model can also be distributed into the client-side to transfer each chunk feature into the chunk-context aware features. And the cloud service provider can also generate variance models for different data scenarios.

Our design's essence is to embed the chunk-context information combined with the chunk's internal structure into its feature. Precisely, if the target chunk is similar to another chunk, the surrounding chunks of the target one may also resemble chunks with a high probability. Conversely,  if the surrounding chunks of a target chunk are similar to other chunks, the target chunk may also be similar. Moreover, there is a similar internal structure among resemble chunks. we leverage the chunk context for resemblance detection which is ignored by the existing schemes.

Besides, there are other benefits from embedding chunk-context into resemblance detection. First, the chunk feature is not easily affected by the modification patterns. These kinds of features are not only related to chunk content but also the chunk-context information.  Second, this design supports the parallel computation for the model training, which is detailed in Subsection \ref{chunk_based context-aware model}. Third, the design can effectively avoid the dimension disaster. D is a fixed pre-defined dimension parameter that could be adjusted based on real applications in our design.


\subsection{N-Sub-chunk Shingles based Initial Feature Extraction Scheme}
\label{initial_feature_extraction_scheme}

In this subsection,  we describe the detail of the n-sub-chunk shingles based initial feature extraction scheme. This scheme aims to generate the initial feature by only considering the chunk content itself for the further embedding chunk-context feature. If two chunks are similar, there always has a common content sequence in each other chunks. Thus, our scheme divides the chunk into many sub-chunks and extracts each sub-chunk's local sensitive hash. The local sensitive hash sequence in a chunk reflects the content structure, denoted as the initial chunk feature. It can avoid the content structure information loss caused by the traditional feature extraction scheme.

The detail of n-sub-chunk shingles based initial feature extraction scheme is shown in Algorithm\ref{n-gram}. There are four input parameters, where $ck_i$ denotes the i-th chunk content, $K$ denotes the number of the sub-words in the $ck_i$, M denotes the dimension of the vector dimension of the $ck_i$, and the $Hash\_Sets$ denotes M hash functions. Firstly, our scheme divides the chunk into $K$ sub-chunks with a fixed size. Secondly, each sub-chunk's local sensitive (LSH) hash value is recorded in a hash array. These LSH hash values are treated as the basic unit in constructing the initial vector. Thirdly, $S$ is a set of unique shingles, each of which is composed of contiguous sub-sequences of sub-chunk in $ck_i$.  These unique shingles embed the content structure by maintaining the sub-sequences of sub-chunk.  Then, each unique shingle is converted into a $sub\_vector$ with M hash functions in $Hash\_Sets$. Finally, the initial vector of $ck_i$, $vector_i$, is the average of these sub\_vectors.



\begin{algorithm}[h]
	\caption{N-sub-chunk shingles Based Initial Feature Extraction Scheme}
	\begin{algorithmic}[1]
		\Require
		$ck_i$:chunk content;
		$K$:number of the sub-words;
		$M$:dimension of the vector;
		$Hash\_Sets$: M hash functions
		\Ensure
		M-dimension rough features, $vector_i$
		\State Split the $ck_i$ into $K$ sub-chunks 
		\For{j=0 to K}
		\State
		hash[j]$\leftarrow$LSH(j-th sub-chunk)
		\EndFor
		\For{r=1 to N}
		\For{j=0 to K}
		\State Set S $\leftarrow$ a string concatenated by the hash[j] and its surrounding r hash values in order
		\EndFor
		\EndFor
		\State s\_size = the element number of S
		\For{g=0 to s\_size}
		\State $e_g$ = g-th element in S
		\State $sub\_vector[g]\leftarrow (hf_0(e_g)$,$hf_1(e_g)$,$\cdots$, $hf_i(e_g))$, where $hf_i$ is the i-th hash function in $Hash\_Sets$
		\EndFor
		\State
		$vector_i$=avg($\frac{sub\_vector[0]}{\Vert sub\_vector[0]\Vert}$,$\cdots$,$\frac{sub\_vector[s\_size]}{\Vert sub\_vector[s\_size]\Vert}$) \\
		\Return  $vector_i$
	\end{algorithmic}
	\label{n-gram}
\end{algorithm}

To represent the chunk content structure in the initial feature, we introduce the sub-chunk and obtain numeric sub-sequences of these sub-chunks. The $ck_i$'s vector representation is associated with these sub-sequences. The essence of the N-sub-chunk shingles-based initial feature extraction scheme is that the contiguous sub-sequence set keeps the same sequence with the chunk content structure. To achieve each dimension value of an initial feature, we randomly transform each shingle into various hash values and put them together in order. It effectively embeds the chunk content information into a feature. To further improve the resemblance detection accuracy, we normalize each dimension value. These initial features are the basement for further embedding the chunk-context information in the chunk-based context-aware model, which is detailed in Subsection\ref{chunk_based context-aware model}.


\subsection{BP-Neural Network-based Chunk-Context Aware Model}
\label{chunk_based context-aware model}

Although the N-sub-chunk shingles based initial feature extraction scheme keeps the chunk content structure into the feature, these features lack chunk-context information. Based on the analysis in Section\ref{Problem_Statement},  two similar chunks may also have similar chunk-context and vice versa. The chunk-context information can improve the quality of the chunk feature. Thus, we further propose a BP-neural network based chunk-context aware model. It embeds the chunk-context information with the chunk content itself to support the resemblance detection in different sizes and make our scheme more robust under various modification patterns.

\begin{figure}[htbp]
	\centering
	\includegraphics[height=5.6cm,width=8.5cm]{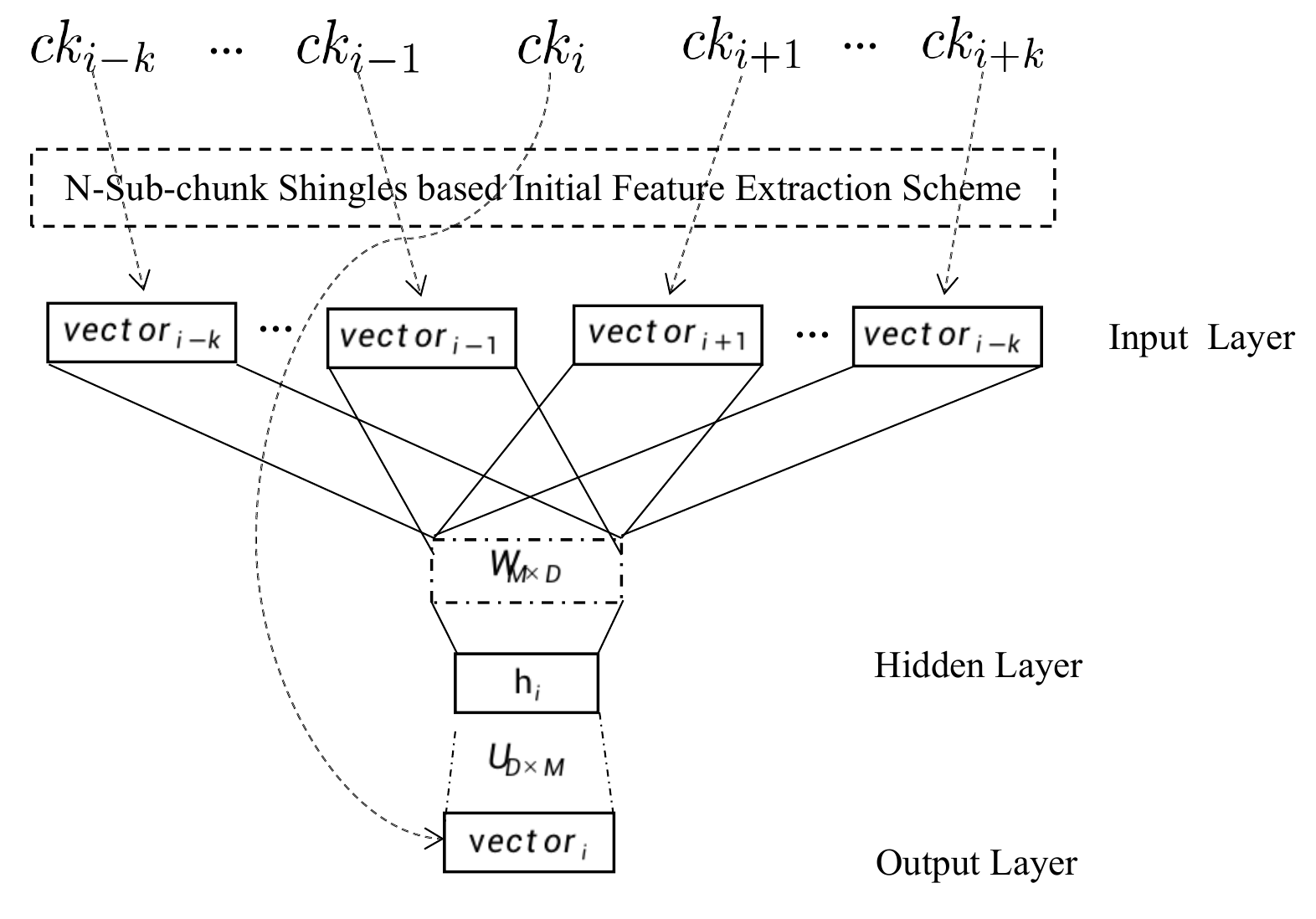}
	\caption{BP-Neural Network-based Chunk-Context Aware Model}
	\label{Model}
\end{figure}

As shown in Figure \ref{Model}, we introduce the BP-Neural Network technology to embed the chunk-context information into the initial feature, which is denoted as the BP-Neural Network-based chunk-context aware model. The initial features are treated as the training data for the BP-Neural network. Then, it calculates a mapping relationship from the surrounding co-occurring chunks, according to the target chunk $ck_i$ of the first k and the last k chunks, based on the training dataset. Specifically, the surrounding chunk initial features of $ck_i$ are the inputs of the model. The $ck_i$'s initial feature is treated as the label of the model output. The hidden layer in the BP-Neural Network-based chunk-context aware model is a vector,$h_i$, with a D dimension. 

\begin{equation}
	\begin{aligned}
		h_i&=W_{M\times D}\cdot(\sum\limits_{i=1}^2Kvector_i)\cdot\frac{1}{2K}
		\label{formula_1}
	\end{aligned}
\end{equation}

\begin{equation}
	\begin{aligned}
		vertor_i&=\frac{1}{2K}\cdot U_{D\times M}\cdot h_i
		\label{formula_2}
	\end{aligned}
\end{equation}

The model is adjusted by the forward feedback and backward feedback mechanism in BP-Neural Network in the training process.  The loss function during the training process is the hierarchical softmax function \cite{DBLP:conf/aaai/PengLSL17}. Since there are large volume data in cloud deduplication scenario under distinct background, cloud service providers can also generate specific training models for different background datasets.  Furthermore, the optimization mapping relationship from the surrounding co-occurred chunks to the target chunk is a chunk-context aware model including two weight matrix, $W_{M\times D}$ and $U_{D\times M}$. The $h_i$ is a hidden layer vector which is calculated based on the Formula\ref{formula_1}.  Moreover, the $h_i$ is a vector of $ck_i$ embedding the chunk-context and chunk internal structure information. And the mapping relationship between chunk's initial feature and the chunk-context feature is based on Formulation\ref{formula_2}. In the predicting process, each new chunk's initial vector, $vector_j$, maps to the context-aware chunk feature, $vector^{'}_{j}$, based on the model as Formula\ref{formula_3}.

\begin{equation}
	\begin{aligned}
		vector^{'}_{i}&=2K\cdot U^{-1}_{D\times M}\cdot vector_i
		\label{formula_3}
	\end{aligned}
\end{equation}

It is noted that the model can be trained and predicted in parallel mode. Specifically, most operations in our model are based on matrix multiplication and addition. Moreover, we defined the chunk-context as the surrounding co-occurring 2K chunks. It significantly avoids the calculation dependency in training, and the multiple CPUs implementation can achieve an optimal convergence speed. What's more, since the model training process can be completed offline, it does not impact the user's experience for cloud storage service. The model storage cost can be decreased by model compression and acceleration, which is not the focus of this paper.



\section{Experimental Evaluations}
\label{experimental_evaluation}

\subsection{Methodology}
\textbf{Implementation and Experimental Setup} We implement the CARD prototype in python. All the experiments are conducted on a clustered platform. The cloud is simulated by three machines. Each machine is equipped with an Intel(R) Core(TM) i7-4790 @3.60Ghz 8 core CPU, 16GB RAM, a 500GB 5400rpm hard disk and is connected to a 100Mbps network. The OS was installed in Ubuntu 18.10 LTS 64-bit System.

\textbf{DataSets} To evaluate the effectiveness of our design, we choose three distinct real-world datasets. The first one is the vmdk file trace, which was collected from initial full backups and subsequent incremental backups
of 6 members of a university research group and was reported by \cite{7208297}. The second one is a SQL dump workload including different periods backup and was reported by \cite{DBLP:conf/msr/BellerGZ17a}. The third one is the Linux Kernel, reported by \cite{DBLP:conf/hpcc/TianLXX18}.  It is noted that these three datasets containing version data for different periods. Distinct version data includes many modification patterns in daily cloud storage usages. Thus, to justify whether the chunk-context information makes the chunk feature robust than other state-of-the-art work, we conduct serials experiments based on these three datasets.

\textbf{Metrics and Configuration} We evaluate the CARD's resemblance detection in two metrics: Delta Compression Ratio (DCR) and the overall time cost for resemblance detection in deduplication. The DCR metric is reported in \cite{DBLP:conf/fast/ZhangX000W19}, which is measured by $\frac{total\quad size\quad  before\quad  delta \quad  compression}{total\quad  size\quad  after\quad  delta\quad  compression}$. The DCR reflects the total space saved by resemblance detection and delta compression. Besides, the overall time cost for resemblance detection is critical for the cloud storage provider, closely related to the user experience for cloud storage service. We evaluate the performance of CARD compared with the state-of-the-art resemblance detection work, N-transform and Finesse\cite{DBLP:conf/fast/ZhangX000W19}. All the systems take the FastCDC \cite{DBLP:conf/usenix/XiaZJFHHLZ16} as the chunking algorithm and the classic Xdelta as the delta compression method \cite{DBLP:conf/usenix/KulkarniDLT04}. We mainly measure the performance of multiple workloads in different metrics.

\subsection{Numerical Results and Discussions}

To evaluate the effectiveness by embedding the chunk-context into features for resemblance detection, we first conduct the experiments on N-transform, Finesse, and CARD with various average chunk size under different workloads. It is because the average chunk size is a critical parameter in deduplication.  We select the most commonly used average chunk size as the variance from 16KB to 512KB. Then, we collect the DCR and the overall time cost for each experiment under a fixed chunk-context feature dimension. It is noted that the chunk-context feature dimension is related to the maximum storage size of cloud storage. To simulate the real cloud storage, we fixed the chunk-context feature dimension in 50, which means the cloud storage could store 1PB data.  

\begin{figure}[htbp]
	\centering
	\includegraphics[height=5cm,width=8cm]{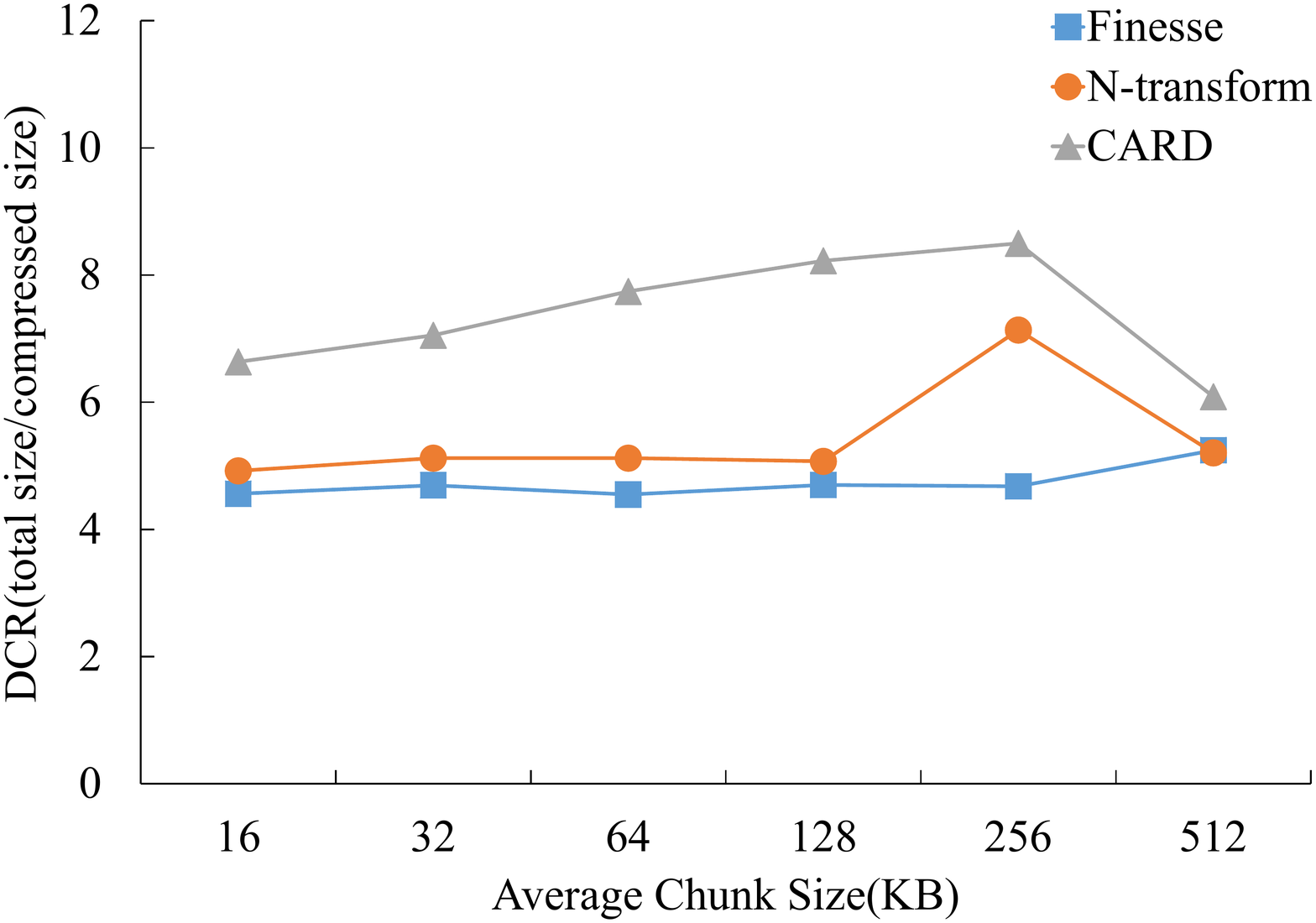}
	\caption{The DCR in SQL Dump Workload}
	\label{fig_SQL_acuuracy}
\end{figure}

As shown in Figure \ref{fig_SQL_acuuracy},  the Finesse has a lower DCR than the N-transform in this workload. When the average chunk size is 16KB, the DCR in N-transform is 0.36 higher than that of Finesse. As the average chunk size increasing, the DCR in Finesse and N-transform is growing slowly. For example, when the average chunk size equals 128KB, the DCR of N-transform and Finesse is 3.02\% and 3.03\% higher than that of 16KB, respectively.  The feature extraction method easily suffers from the modification discussed in Section\ref{Problem_Statement}, while the N-transform's feature extraction is more robust than the Finesse.

Fortunately, our method achieves a better DCR than N-transform and Finesse in the SQL dump workload, as shown in Figure \ref{fig_SQL_acuuracy}. Compared with the N-transform and Finesse, our method improves the DCR by 45.49\% and 51.19\% when the average chunk size is 16KB,  respectively. As the chunk average size is 128KB, the CARD is 62.02\% and 75.03\% higher than the N-transform and Finesse in DCR. It does not contradict the view that the smaller the chunk size is, the more redundancy can be removed in deduplication. There are lots of redundancy in the SQL dump workload. The same content in a small average chunk size generates many delta files. In contrast, small delta files are generated in a large average chunk size. The previous one costs more storage than the latter. Thus, although it provides qualified chunk-context information in a small average chunk size, there still has a DCR improving space with the average chunk size increasing. As the average chunk size is 512KB, Both CARD and N-transform's DCR decrease because too sizeable average chunk size deteriorates the similarity detection effectiveness. Nevertheless, our method is still 16.05\% higher than finesse's DCR.



\begin{figure}[htbp]
	\centering
	\includegraphics[height=5cm,width=8cm]{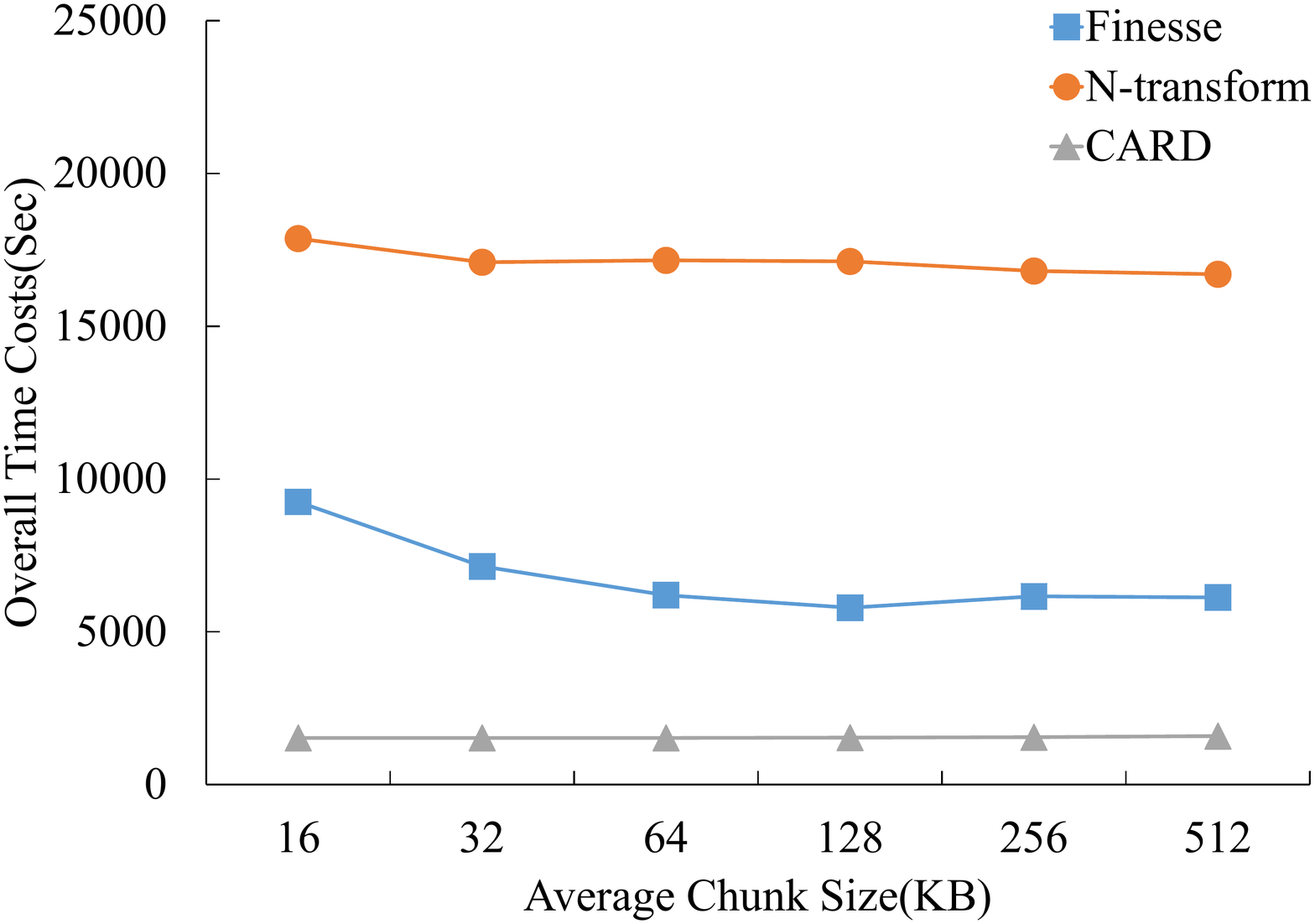}
	\caption{The Overall Time Cost in SQL Dump Workload}
	\label{fig_SQL_performance}
\end{figure}

Besides, we also statistic the overall time cost for Finesse, N-transform, and CARD in the SQL dump workload, as shown in Figure \ref{fig_SQL_performance}. When the average chunk size is 16KB, the CARD is 6$\times$ and  11$\times$ faster than the Finesse and N-transform, respectively. As the average chunk size increasing, the overall time of N-transform and CARD fluctuate slightly in a fixed range. N-transform's primary time consumption is the Rabin fingerprint value and the linear transformation calculations, which are independent of the average chunk size. Moreover, the chunk-context feature is generated through the N-sub-chunk shingles based initial feature extraction and the chunk-context embedding process. The previous initial feature extraction is related to the feature dimension, but not to the average chunk size. However, Finesse's feature sorting costs too much when the average chunk size is small, which dramatically raises the sorting number as the chunk number increases. Nevertheless, the CARD is still 3.8$\times$ than Finesse when the average chunk size is 512KB.

\begin{figure}[htbp]
	\centering
	\includegraphics[height=5cm,width=8cm]{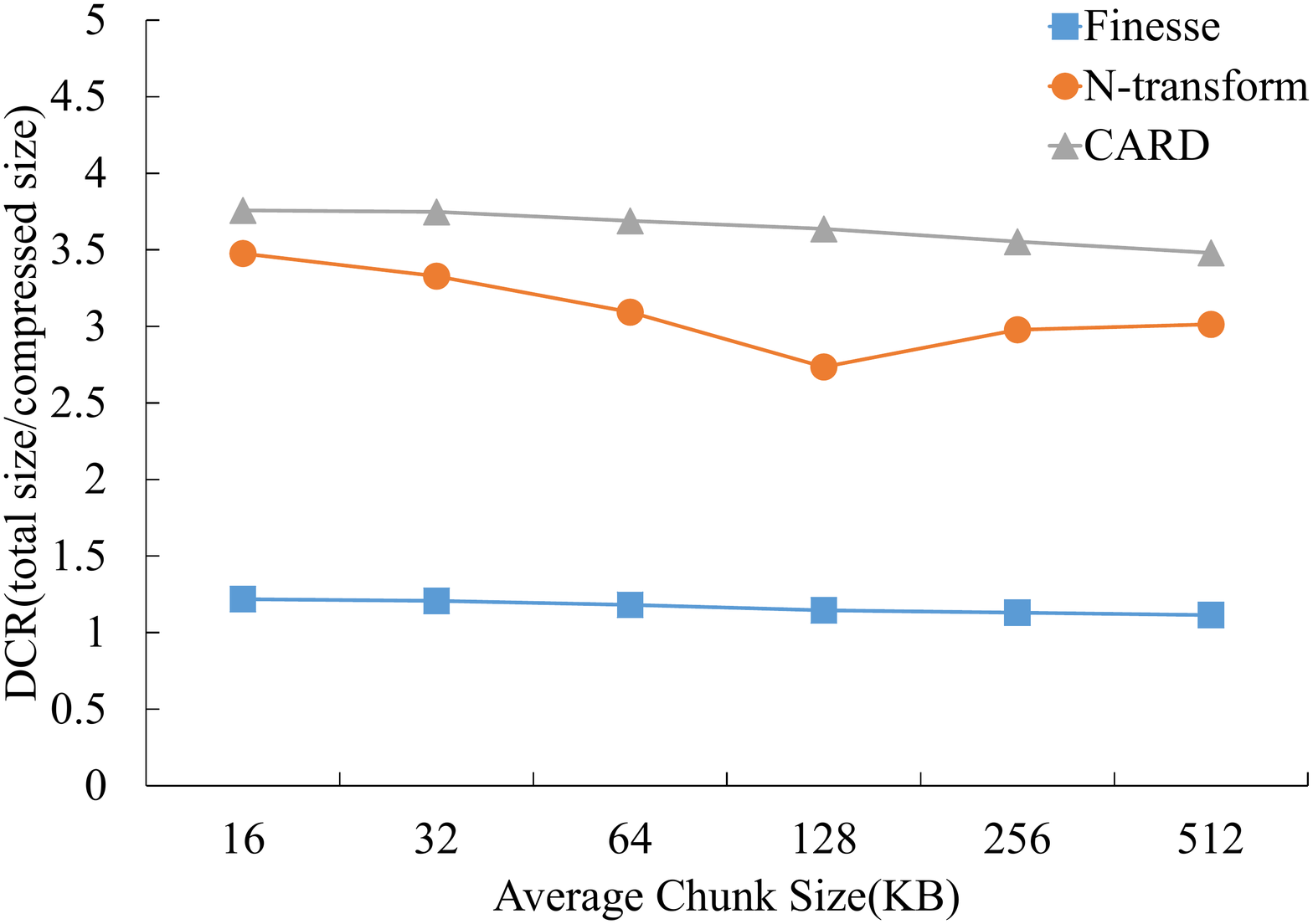}
	\caption{The DCR in VMDK Workload}
	\label{fig_VMDK_acuuracy}
\end{figure}

Similarly, we conduct the same configuration in VMDK and Linux Kernal workloads.  As shown in Figure \ref{fig_VMDK_acuuracy}, the DCR of Finesse, N-transform, and CARD is decreasing slowly with the increasing average chunk size. Specifically, the DCR of CARD, N-transform SF and Finesse, is reduced by 7.37\%, 13.31\%, and 8.54\% from 16KB to 512KB average chunk size, respectively. The reason is that the data modification pattern in VMDK tends to be random compared with the SQL dump workload. The larger average chunk deteriorates the effectiveness of the features in all methods. Thus, it does not occur the DCR increasing tendency when the average chunk size increases. Nevertheless, the DCR of CARD is higher than the Finesse and N-transform with variance average chunk size.


In Linux Kernel workload, we also conduct the experiments by changing the average chunk size from 16KB to 512 KB, although a large number of the file in this workload is less than 4KB except for some resource files. Our goal is to test the effectiveness of various resemblance methods under this extreme situation. As shown in Figure \ref{fig_VMDK_acuuracy}, our method has outperformed the Finesse. For example, the DCR of our method is $2.25\times$ higher than that of Finesse when the average chunk size is 512KB. However, the CARD and N-transform have similar DCR under these kinds of extreme conditions. It is because most of the files in Linux Kernel become a single chunk without the chunk-context information under these extreme conditions. Thus, CARD's chunk-context feature will degenerate into a feature only related to the chunk content internal structures.

\begin{figure}[htbp]
	\centering
	\includegraphics[height=5cm,width=8cm]{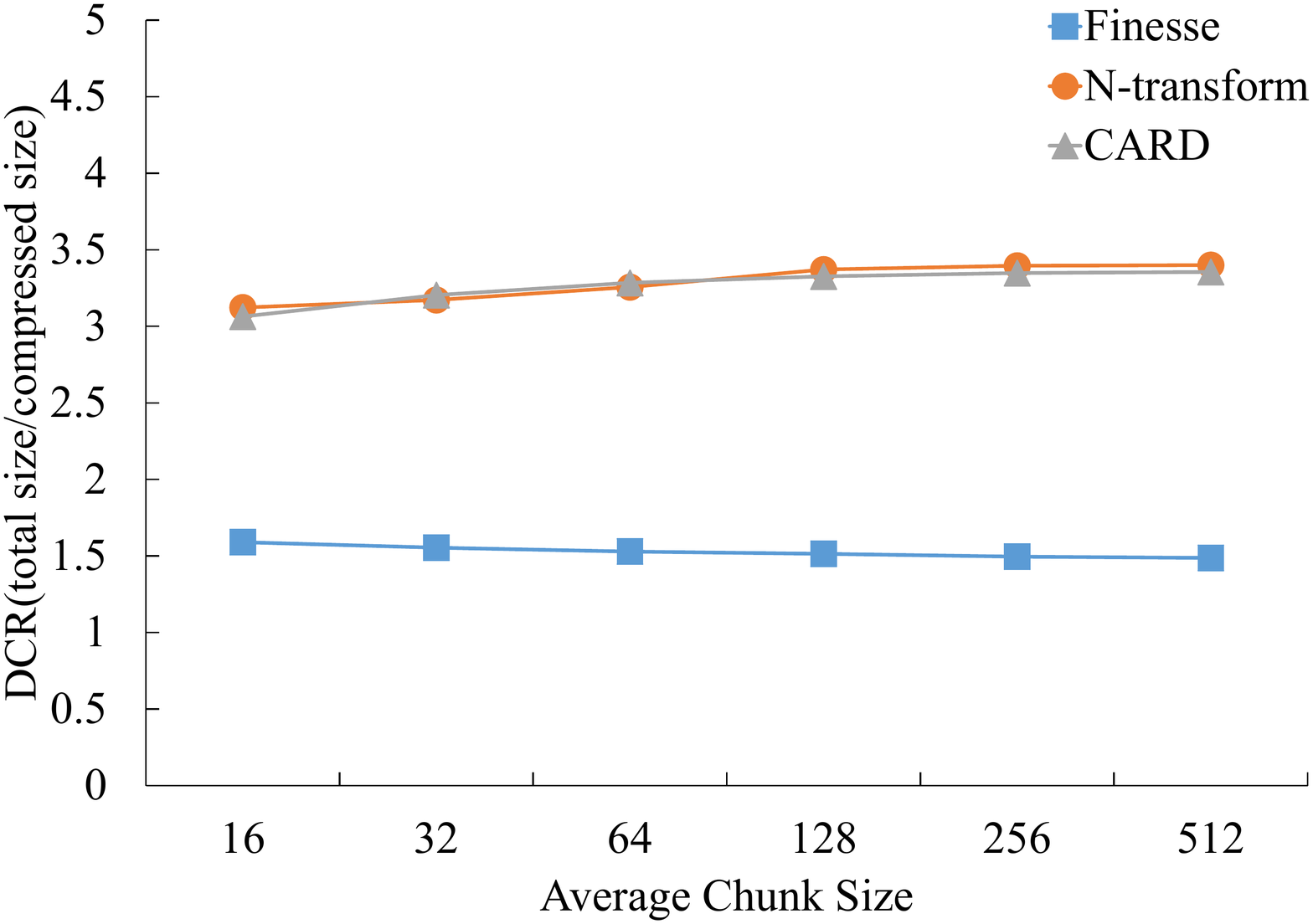}
	\caption{The DCR in Linux Kernel Workload}
	\label{fig_kernel_accuracy}
\end{figure}


Besides, we also compare the overall time cost under VMDK and Linux Kernel workloads. In the VMDK dataset, when the average chunk size is 256KB, CARD is faster than Finesse and N-transform SF by 5.6$\times$ and 17.88$\times$, respectively. Nevertheless, when the average chunk size is 16KB, Finesse is slower than N-transform by 3.2$\times$. Moreover, as the average chunk size increases, Finesse is finally stabilizes and is 2.2$\times$ faster than N-transform. It is because each file generates more chunks when the average chunk size is less than 32KB. Therefore, the sorting process in Finesse dramatically increases. Thus, the sorting cost is much more than the time cost for extracting features. Once the average chunk size is more than 32KB, the cost of extracting feature becomes the dominant time consumption.

\begin{figure}[htbp]
	\centering
	\includegraphics[height=5cm,width=8cm]{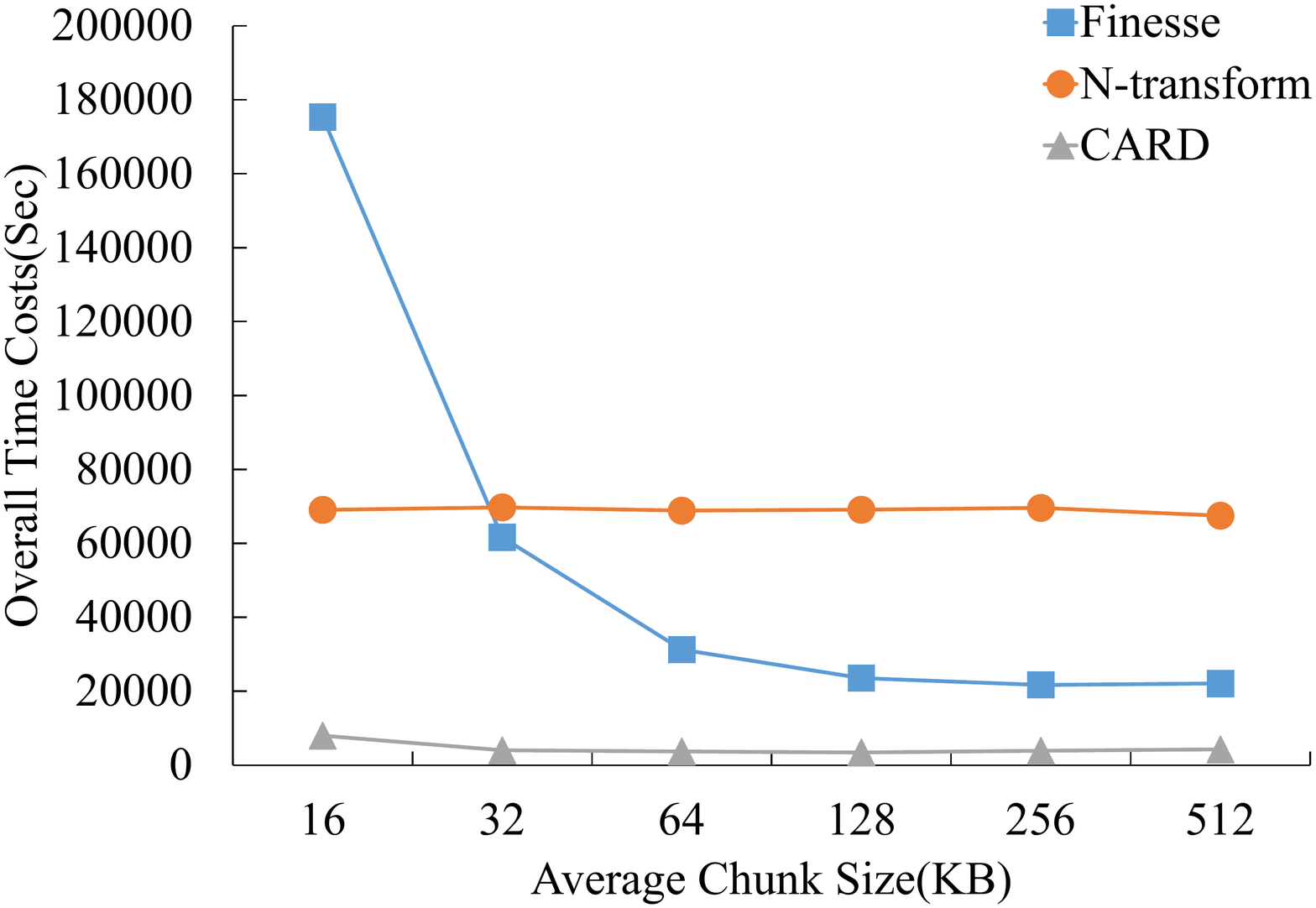}
	\caption{The Overall Time Cost in VMDK Workload}
	\label{fig_VMDK_performance}
\end{figure}

\begin{figure}[htbp]
	\centering
	\includegraphics[height=5cm,width=8cm]{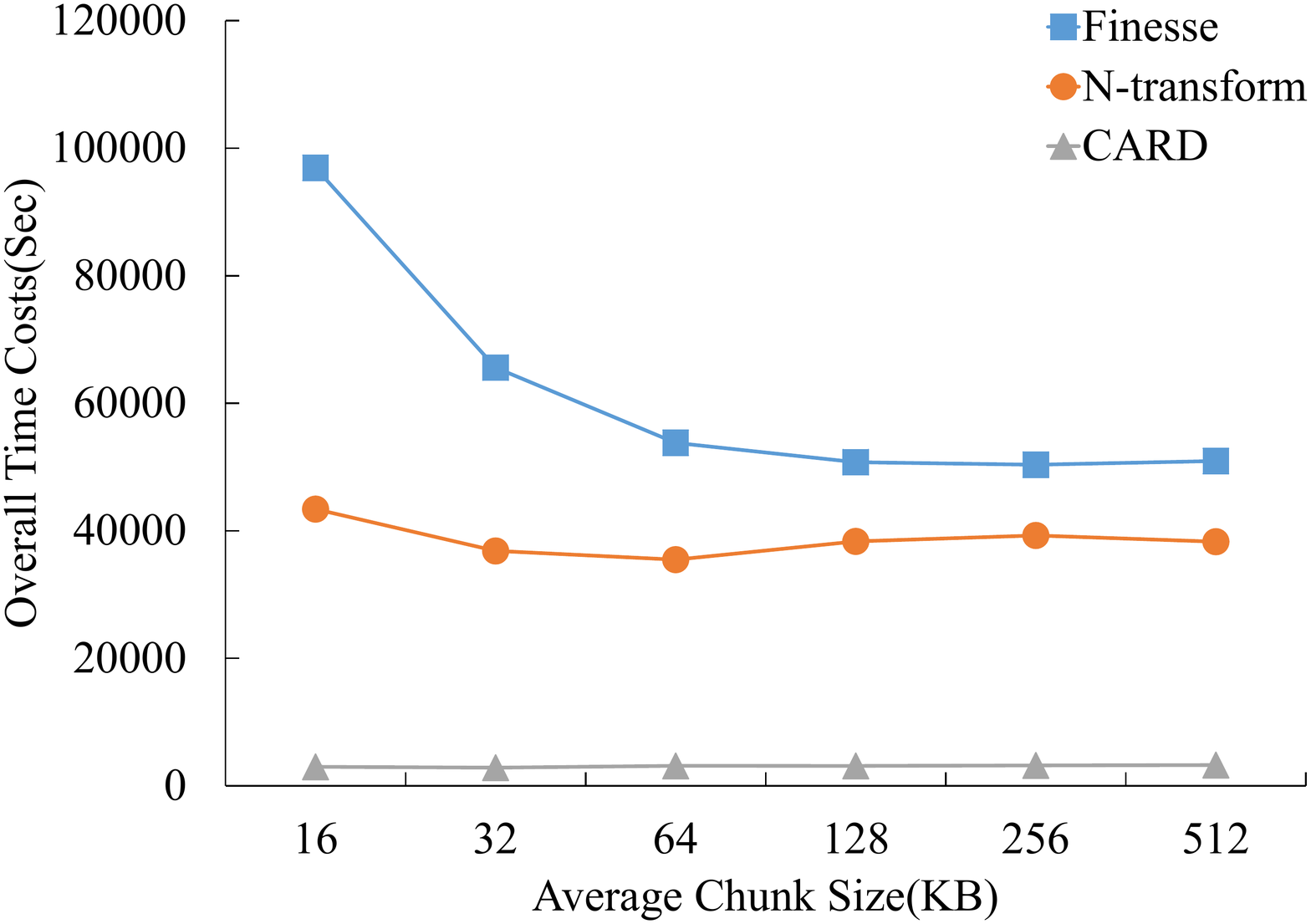}
	\caption{The Overall Time Cost in Linux Kernel Workload}
	\label{fig_kernel_performance}
\end{figure}

As shown in Figure\ref{fig_kernel_performance}, the overall time cost CARD is far less than that of Finesse and N-transform when the average chunk size is changed from 16KB to 512 KB. Specifically, when the average chunk size is 16KB, the Finesse time consumption is 32 times longer than our method's overall time cost. Although the CARD has a similar DCR with  N-transform under these extreme conditions, most of the feature extraction in CARD are hash calculation without the Rabin fingerprint value's linear transform in N-transform or the Rabin fingerprint value's sorting in Finesse. Thus, the overall time cost under small average chunks size in Finesse and N-transform is much higher than the time consumption under larger average chunks size.


\begin{table}[htbp]
    \centering
	\caption{ Overall Cost TImes and DCR of CARD with Distinct Deature Dimensions}
	\label{talbe_experiemnt_distinct_dimensions}
	\vspace{0.2cm}
	\begin{tabular}{|p{1.85cm}|p{1.5cm}|l|l|} 
		
		\hline 
		\emph{Dataset}&\emph{Dimension}&\emph{Time(s)}&\emph{DCR} \\ \hline \hline
		
		\multirow{5}*{SQL dump}
		&\centering 40&1514.86&7.73\\
		\cline{2-4}
		&\centering 50&1527.14&7.74${\color{red}(+0.13\%)}$\\
		\cline{2-4}
		&\centering 60&1693.37&7.79${\color{red}(+0.78\%)}$\\
		\cline{2-4}
		&\centering 70&1671.33&7.77${\color{red}(+0.52\%)}$\\
		\cline{2-4}
		&\centering 80&1715.07&8.01${\color{red}(+3.62\%)}$\\ \hline \hline
		
		\multirow{5}*{Linux Kernel}
		&\centering 40&3073.52&3.29\\
		\cline{2-4}
		&\centering 50&3128.84&3.28${\color{blue}(-0.30\%)}$\\
		\cline{2-4}
		&\centering 60&3575.78&3.29${\color{black}(0.00\%)}$\\
		\cline{2-4}
		&\centering 70&3970.23&3.29${\color{black}(0.00\%)}$\\
		\cline{2-4}
		&\centering 80&4344.69&3.32${\color{red}(+0.91\%)}$\\ \hline \hline
		
		\multirow{5}*{VMDK}
		&\centering 40&3681.18&3.69\\
		\cline{2-4}
		&\centering 50&3719.48&3.69${\color{black}(0.00\%)}$\\
		\cline{2-4}
		&\centering 60&3726.99&3.69${\color{black}(0.00\%)}$\\
		\cline{2-4}
		&\centering 70&3759.93&3.69${\color{black}(0.00\%)}$\\
		\cline{2-4}
		&\centering 80&3824.97&3.69${\color{black}(0.00\%)}$\\ \hline 
	\end{tabular}
\end{table}

Finally, we evaluate the CARD performance on distinct workloads with various feature dimension settings in fixed average chunk size. As we discussed before, the feature dimension in 50 denotes that the cloud storage could store 1PB data. To simulate the real cloud storage, we changed the feature dimension from 40 to 80. Once the feature dimension is 80, it means the maximum volume of cloud storage is $2^{10}$ZB. In Table\ref{talbe_experiemnt_distinct_dimensions}, the overall time cost is gradually increasing as the feature dimension expanding under distinct workloads. Moreover, the CARD DCR is slightly increasing when the dimension is changed from 40 to 80. In Linux Kernel and VMDK workloads, the DCR of our method is almost keeps the same value when the dimension is changed from 40 to 80.  It is because that some modification patterns in VMDK and Linux Kernel are tend to be random which is no conductive to the chunk-context embedding, which is a problem we will further solve in the future.  Nevertheless, our method is still outperforms than the state-of-the-art resemblance work, such as N-transform and Finesse, in distinct workloads.


\section{Conclusions}
\label{conclusion}
The paper analyzes the state-of-the-art resemblance detection technology in deduplication and presents its limitations in ignoring the chunk-context information. Based on our observation, if the chunk feature in resemblance detection technology is only related to the chunk content itself, it easily suffers from the various modifications. Moreover, if the two chunks surrounding chunks are resemblance, these two chunks might be similar with high probability and vice versa. Inspired by the phenomenon, we propose a chunk-context aware resemblance detection, called CARD, to decrease various modification impacts. In CARD, the N-sub-chunk shingles based initial feature extraction is proposed to extract a preliminary feature by considering the chunk content internal structure. Furthermore, a BP-Neural network-based chunk-context aware model is also proposed to embed the chunk-context information based on the preliminary features. Finally, based on the conducted experimental results, CARD outperforms the state-of-the-art resemblance detection solutions in deduplication, such as N-transform and Finesse. It can remove up to 75.03\% more redundancy data and accelerate the resemblance detection by 5.6$\times\sim$17.8$\times$ compared with the state-of-the-art resemblance detection work, N-transform and Finesse.

\bibliographystyle{unsrtnat}
\bibliography{template}  

\begin{thebibliography}{33}
\providecommand{\natexlab}[1]{#1}
\providecommand{\url}[1]{\texttt{#1}}
\expandafter\ifx\csname urlstyle\endcsname\relax
  \providecommand{\doi}[1]{doi: #1}\else
  \providecommand{\doi}{doi: \begingroup \urlstyle{rm}\Url}\fi

\bibitem[Sharma et~al.(2020)Sharma, Jindal, and Malaya]{1}
Pratima Sharma, Rajni Jindal, and Dutta~Borah Malaya.
\newblock Blockchain technology for cloud storage: {A} systematic literature
  review.
\newblock \emph{{ACM} Comput. Surv.}, 53\penalty0 (4):\penalty0 89:1--89:32,
  2020.
\newblock \doi{10.1145/3403954}.
\newblock URL \url{https://doi.org/10.1145/3403954}.

\bibitem[Chaudhuri et~al.(2007)Chaudhuri, Sarma, Ganti, and Kaushik]{2}
Surajit Chaudhuri, Anish~Das Sarma, Venkatesh Ganti, and Raghav Kaushik.
\newblock Leveraging aggregate constraints for deduplication.
\newblock In Chee~Yong Chan, Beng~Chin Ooi, and Aoying Zhou, editors,
  \emph{Proceedings of the {ACM} {SIGMOD} International Conference on
  Management of Data, Beijing, China, June 12-14, 2007}, pages 437--448. {ACM},
  2007.
\newblock \doi{10.1145/1247480.1247530}.
\newblock URL \url{https://doi.org/10.1145/1247480.1247530}.

\bibitem[Meyer and Bolosky(2011)]{3}
Dutch~T. Meyer and William~J. Bolosky.
\newblock A study of practical deduplication.
\newblock In Gregory~R. Ganger and John Wilkes, editors, \emph{9th {USENIX}
  Conference on File and Storage Technologies, San Jose, CA, USA, February
  15-17, 2011}, pages 1--13. {USENIX}, 2011.
\newblock URL
  \url{http://www.usenix.org/events/fast11/tech/techAbstracts.html\#Meyer}.

\bibitem[Tian et~al.(2018)Tian, Li, Xiao, and Xu]{DBLP:conf/hpcc/TianLXX18}
Wenlong Tian, Ruixuan Li, Weijun Xiao, and Zhiyong Xu.
\newblock Pts-dep: {A} high-performance two-party secure deduplication for
  cloud storage.
\newblock In \emph{20th {IEEE} International Conference on High Performance
  Computing and Communications; 16th {IEEE} International Conference on Smart
  City; 4th {IEEE} International Conference on Data Science and Systems,
  HPCC/SmartCity/DSS 2018, Exeter, United Kingdom, June 28-30, 2018}, pages
  700--707. {IEEE}, 2018.
\newblock \doi{10.1109/HPCC/SmartCity/DSS.2018.00122}.
\newblock URL \url{https://doi.org/10.1109/HPCC/SmartCity/DSS.2018.00122}.

\bibitem[Muthitacharoen et~al.(2001)Muthitacharoen, Chen, and
  Mazi{\`{e}}res]{DBLP:conf/sosp/MuthitacharoenCM01}
Athicha Muthitacharoen, Benjie Chen, and David Mazi{\`{e}}res.
\newblock A low-bandwidth network file system.
\newblock In Keith Marzullo and Mahadev Satyanarayanan, editors,
  \emph{Proceedings of the 18th {ACM} Symposium on Operating System Principles,
  {SOSP} 2001, Chateau Lake Louise, Banff, Alberta, Canada, October 21-24,
  2001}, pages 174--187. {ACM}, 2001.
\newblock \doi{10.1145/502034.502052}.
\newblock URL \url{https://doi.org/10.1145/502034.502052}.

\bibitem[Eshghi and Tang(2005)]{eshghi2005framework}
Kave Eshghi and Hsiu~Khuern Tang.
\newblock A framework for analyzing and improving content-based chunking
  algorithms.
\newblock \emph{Hewlett-Packard Labs Technical Report TR}, 30\penalty0 (2005),
  2005.

\bibitem[Tian et~al.(2017)Tian, Li, Xu, and Xiao]{DBLP:conf/ipccc/TianLXX17}
Wenlong Tian, Ruixuan Li, Zhiyong Xu, and Weijun Xiao.
\newblock Does the content defined chunking really solve the local boundary
  shift problem?
\newblock In \emph{36th {IEEE} International Performance Computing and
  Communications Conference, {IPCCC} 2017, San Diego, CA, USA, December 10-12,
  2017}, pages 1--8. {IEEE} Computer Society, 2017.
\newblock \doi{10.1109/PCCC.2017.8280445}.
\newblock URL \url{https://doi.org/10.1109/PCCC.2017.8280445}.

\bibitem[Xia et~al.(2016)Xia, Zhou, Jiang, Feng, Hua, Hu, Liu, and
  Zhang]{DBLP:conf/usenix/XiaZJFHHLZ16}
Wen Xia, Yukun Zhou, Hong Jiang, Dan Feng, Yu~Hua, Yuchong Hu, Qing Liu, and
  Yucheng Zhang.
\newblock Fastcdc: a fast and efficient content-defined chunking approach for
  data deduplication.
\newblock In Ajay Gulati and Hakim Weatherspoon, editors, \emph{2016 {USENIX}
  Annual Technical Conference, {USENIX} {ATC} 2016, Denver, CO, USA, June
  22-24, 2016}, pages 101--114. {USENIX} Association, 2016.
\newblock URL
  \url{https://www.usenix.org/conference/atc16/technical-sessions/presentation/xia}.

\bibitem[Aronovich et~al.(2009)Aronovich, Asher, Bachmat, Bitner, Hirsch, and
  Klein]{DBLP:conf/systor/AronovichABBHK09}
Lior Aronovich, Ron Asher, Eitan Bachmat, Haim Bitner, Michael Hirsch, and
  Shmuel~T. Klein.
\newblock The design of a similarity based deduplication system.
\newblock In Miriam Allalouf, Michael Factor, and Dror~G. Feitelson, editors,
  \emph{Proceedings of of {SYSTOR} 2009: The Israeli Experimental Systems
  Conference 2009, Haifa, Israel, May 4-6, 2009}, {ACM} International
  Conference Proceeding Series, page~6. {ACM}, 2009.
\newblock \doi{10.1145/1534530.1534539}.
\newblock URL \url{https://doi.org/10.1145/1534530.1534539}.

\bibitem[Douglis and Iyengar(2003)]{DBLP:conf/usenix/DouglisI03}
Fred Douglis and Arun Iyengar.
\newblock Application-specific delta-encoding via resemblance detection.
\newblock In \emph{Proceedings of the General Track: 2003 {USENIX} Annual
  Technical Conference, June 9-14, 2003, San Antonio, Texas, {USA}}, pages
  113--126. {USENIX}, 2003.
\newblock URL \url{http://www.usenix.org/events/usenix03/tech/douglis.html}.

\bibitem[Forman et~al.(2005)Forman, Eshghi, and
  Chiocchetti]{DBLP:conf/kdd/FormanEC05}
George Forman, Kave Eshghi, and Stephane Chiocchetti.
\newblock Finding similar files in large document repositories.
\newblock In Robert Grossman, Roberto~J. Bayardo, and Kristin~P. Bennett,
  editors, \emph{Proceedings of the Eleventh {ACM} {SIGKDD} International
  Conference on Knowledge Discovery and Data Mining, Chicago, Illinois, USA,
  August 21-24, 2005}, pages 394--400. {ACM}, 2005.
\newblock \doi{10.1145/1081870.1081916}.
\newblock URL \url{https://doi.org/10.1145/1081870.1081916}.

\bibitem[Pucha et~al.(2007)Pucha, Andersen, and
  Kaminsky]{DBLP:conf/nsdi/PuchaAK07}
Himabindu Pucha, David~G. Andersen, and Michael Kaminsky.
\newblock Exploiting similarity for multi-source downloads using file
  handprints.
\newblock In Hari Balakrishnan and Peter Druschel, editors, \emph{4th Symposium
  on Networked Systems Design and Implementation {(NSDI} 2007), April 11-13,
  2007, Cambridge, Massachusetts, USA, Proceedings}. {USENIX}, 2007.
\newblock URL \url{http://www.usenix.org/events/nsdi07/tech/pucha.html}.

\bibitem[Xu et~al.(2017)Xu, Pavlo, Sengupta, and
  Ganger]{DBLP:conf/sigmod/XuPSG17}
Lianghong Xu, Andrew Pavlo, Sudipta Sengupta, and Gregory~R. Ganger.
\newblock Online deduplication for databases.
\newblock In Semih Salihoglu, Wenchao Zhou, Rada Chirkova, Jun Yang, and Dan
  Suciu, editors, \emph{Proceedings of the 2017 {ACM} International Conference
  on Management of Data, {SIGMOD} Conference 2017, Chicago, IL, USA, May 14-19,
  2017}, pages 1355--1368. {ACM}, 2017.
\newblock \doi{10.1145/3035918.3035938}.
\newblock URL \url{https://doi.org/10.1145/3035918.3035938}.

\bibitem[Kulkarni et~al.(2004)Kulkarni, Douglis, LaVoie, and
  Tracey]{DBLP:conf/usenix/KulkarniDLT04}
Purushottam Kulkarni, Fred Douglis, Jason~D. LaVoie, and John~M. Tracey.
\newblock Redundancy elimination within large collections of files.
\newblock In \emph{Proceedings of the General Track: 2004 {USENIX} Annual
  Technical Conference, June 27 - July 2, 2004, Boston Marriott Copley Place,
  Boston, MA, {USA}}, pages 59--72. {USENIX}, 2004.
\newblock URL
  \url{http://www.usenix.org/publications/library/proceedings/usenix04/tech/general/kulkarni.html}.

\bibitem[Shilane et~al.(2012{\natexlab{a}})Shilane, Wallace, Huang, and
  Hsu]{DBLP:conf/hotstorage/ShilaneWHH12}
Philip Shilane, Grant Wallace, Mark Huang, and Windsor Hsu.
\newblock Delta compressed and deduplicated storage using stream-informed
  locality.
\newblock In Raju Rangaswami, editor, \emph{4th {USENIX} Workshop on Hot Topics
  in Storage and File Systems, HotStorage'12, Boston, MA, USA, June 13-14,
  2012}. {USENIX} Association, 2012{\natexlab{a}}.
\newblock URL
  \url{https://www.usenix.org/conference/hotstorage12/workshop-program/presentation/shilane}.

\bibitem[Rabin(1981)]{rabin1981fingerprinting}
M.O. Rabin.
\newblock \emph{Fingerprinting by Random Polynomials}.
\newblock Center for Research in Computing Technology: Center for Research in
  Computing Technology. Center for Research in Computing Techn., Aiken
  Computation Laboratory, Univ., 1981.
\newblock URL \url{https://books.google.com.sg/books?id=Emu\_tgAACAAJ}.

\bibitem[Zhang et~al.(2019)Zhang, Xia, Feng, Jiang, Hua, and
  Wang]{DBLP:conf/fast/ZhangX000W19}
Yucheng Zhang, Wen Xia, Dan Feng, Hong Jiang, Yu~Hua, and Qiang Wang.
\newblock Finesse: Fine-grained feature locality based fast resemblance
  detection for post-deduplication delta compression.
\newblock In Arif Merchant and Hakim Weatherspoon, editors, \emph{17th {USENIX}
  Conference on File and Storage Technologies, {FAST} 2019, Boston, MA,
  February 25-28, 2019}, pages 121--128. {USENIX} Association, 2019.
\newblock URL
  \url{https://www.usenix.org/conference/fast19/presentation/zhang}.

\bibitem[Clements et~al.(2009)Clements, Ahmad, Vilayannur, and
  Li]{DBLP:conf/usenix/ClementsAVL09}
Austin~T. Clements, Irfan Ahmad, Murali Vilayannur, and Jinyuan Li.
\newblock Decentralized deduplication in {SAN} cluster file systems.
\newblock In Geoffrey~M. Voelker and Alec Wolman, editors, \emph{2009 {USENIX}
  Annual Technical Conference, San Diego, CA, USA, June 14-19, 2009}. {USENIX}
  Association, 2009.
\newblock URL
  \url{https://www.usenix.org/conference/usenix-09/decentralized-deduplication-san-cluster-file-systems}.

\bibitem[Lillibridge et~al.(2013)Lillibridge, Eshghi, and
  Bhagwat]{DBLP:conf/fast/LillibridgeEB13}
Mark Lillibridge, Kave Eshghi, and Deepavali Bhagwat.
\newblock Improving restore speed for backup systems that use inline
  chunk-based deduplication.
\newblock In Keith~A. Smith and Yuanyuan Zhou, editors, \emph{Proceedings of
  the 11th {USENIX} conference on File and Storage Technologies, {FAST} 2013,
  San Jose, CA, USA, February 12-15, 2013}, pages 183--198. {USENIX}, 2013.
\newblock URL
  \url{https://www.usenix.org/conference/fast13/technical-sessions/presentation/lillibridge}.

\bibitem[Lillibridge et~al.(2009)Lillibridge, Eshghi, Bhagwat, Deolalikar,
  Trezis, and Camble]{DBLP:conf/fast/LillibridgeEBDTC09}
Mark Lillibridge, Kave Eshghi, Deepavali Bhagwat, Vinay Deolalikar, Greg
  Trezis, and Peter Camble.
\newblock Sparse indexing: Large scale, inline deduplication using sampling and
  locality.
\newblock In Margo~I. Seltzer and Richard Wheeler, editors, \emph{7th {USENIX}
  Conference on File and Storage Technologies, February 24-27, 2009, San
  Francisco, CA, {USA.} Proceedings}, pages 111--123. {USENIX}, 2009.
\newblock URL
  \url{http://www.usenix.org/events/fast09/tech/full\_papers/lillibridge/lillibridge.pdf}.

\bibitem[Lu et~al.(2012)Lu, Chambliss, Glider, and
  Constantinescu]{DBLP:conf/systor/LuCGC12}
Maohua Lu, David~D. Chambliss, Joseph~S. Glider, and Cornel Constantinescu.
\newblock Insights for data reduction in primary storage: a practical analysis.
\newblock In \emph{The 5th Annual International Systems and Storage Conference,
  {SYSTOR} '12, Haifa, Israel, June 4-6, 2012}, page~17. {ACM}, 2012.
\newblock \doi{10.1145/2367589.2367606}.
\newblock URL \url{https://doi.org/10.1145/2367589.2367606}.

\bibitem[Fu et~al.(2014)Fu, Feng, Hua, He, Chen, Xia, Huang, and
  Liu]{DBLP:conf/usenix/FuFHHCXHL14}
Min Fu, Dan Feng, Yu~Hua, Xubin He, Zuoning Chen, Wen Xia, Fangting Huang, and
  Qing Liu.
\newblock Accelerating restore and garbage collection in deduplication-based
  backup systems via exploiting historical information.
\newblock In Garth Gibson and Nickolai Zeldovich, editors, \emph{2014 {USENIX}
  Annual Technical Conference, {USENIX} {ATC} '14, Philadelphia, PA, USA, June
  19-20, 2014}, pages 181--192. {USENIX} Association, 2014.
\newblock URL
  \url{https://www.usenix.org/conference/atc14/technical-sessions/presentation/fu\_min}.

\bibitem[Upadhyay et~al.(2012)Upadhyay, Balihalli, Ivaturi, and
  Rao]{2012Deduplication}
Amrita Upadhyay, Pratibha~R Balihalli, Shashibhushan Ivaturi, and Shrisha Rao.
\newblock Deduplication and compression techniques in cloud design.
\newblock In \emph{Systems Conference}, 2012.

\bibitem[Oh et~al.(2018)Oh, Park, Yoon, Kim, Lee, Weil, Yeom, and
  Jung]{2018Design}
Myoungwon Oh, Sejin Park, Jungyeon Yoon, Sangjae Kim, Kang~Won Lee, Sage Weil,
  Heon~Y. Yeom, and Myoungsoo Jung.
\newblock Design of global data deduplication for a scale-out distributed
  storage system.
\newblock In \emph{2018 IEEE 38th International Conference on Distributed
  Computing Systems (ICDCS)}, 2018.

\bibitem[Jin and Miller(2009)]{DBLP:conf/systor/JinM09}
Keren Jin and Ethan~L. Miller.
\newblock The effectiveness of deduplication on virtual machine disk images.
\newblock In Miriam Allalouf, Michael Factor, and Dror~G. Feitelson, editors,
  \emph{Proceedings of of {SYSTOR} 2009: The Israeli Experimental Systems
  Conference 2009, Haifa, Israel, May 4-6, 2009}, {ACM} International
  Conference Proceeding Series, page~7. {ACM}, 2009.
\newblock \doi{10.1145/1534530.1534540}.
\newblock URL \url{https://doi.org/10.1145/1534530.1534540}.

\bibitem[Srinivasan et~al.(2012)Srinivasan, Bisson, Goodson, and
  Voruganti]{DBLP:conf/fast/SrinivasanBGV12}
Kiran Srinivasan, Timothy Bisson, Garth~R. Goodson, and Kaladhar Voruganti.
\newblock idedup: latency-aware, inline data deduplication for primary storage.
\newblock In William~J. Bolosky and Jason Flinn, editors, \emph{Proceedings of
  the 10th {USENIX} conference on File and Storage Technologies, {FAST} 2012,
  San Jose, CA, USA, February 14-17, 2012}, page~24. {USENIX} Association,
  2012.
\newblock URL
  \url{https://www.usenix.org/conference/fast12/idedup-latency-aware-inline-data-deduplication-primary-storage}.

\bibitem[Zhao et~al.(2020)Zhao, Albahar, Abraham, Chen, Tarasov, Skourtis,
  Rupprecht, Anwar, and Butt]{DBLP:conf/usenix/ZhaoAACTSRAB20}
Nannan Zhao, Hadeel Albahar, Subil Abraham, Keren Chen, Vasily Tarasov,
  Dimitrios Skourtis, Lukas Rupprecht, Ali Anwar, and Ali~Raza Butt.
\newblock Duphunter: Flexible high-performance deduplication for docker
  registries.
\newblock In Ada Gavrilovska and Erez Zadok, editors, \emph{2020 {USENIX}
  Annual Technical Conference, {USENIX} {ATC} 2020, July 15-17, 2020}, pages
  769--783. {USENIX} Association, 2020.
\newblock URL \url{https://www.usenix.org/conference/atc20/presentation/zhao}.

\bibitem[Shilane et~al.(2012{\natexlab{b}})Shilane, Huang, Wallace, and
  Hsu]{DBLP:conf/fast/ShilaneHWH12}
Philip Shilane, Mark Huang, Grant Wallace, and Windsor Hsu.
\newblock {WAN} optimized replication of backup datasets using stream-informed
  delta compression.
\newblock In William~J. Bolosky and Jason Flinn, editors, \emph{Proceedings of
  the 10th {USENIX} conference on File and Storage Technologies, {FAST} 2012,
  San Jose, CA, USA, February 14-17, 2012}, page~5. {USENIX} Association,
  2012{\natexlab{b}}.
\newblock URL
  \url{https://www.usenix.org/conference/fast12/wan-optimized-replication-backup-datasets-using-stream-informed-delta-compression}.

\bibitem[Zhu et~al.(2008)Zhu, Li, and Patterson]{DBLP:conf/fast/ZhuLP08}
Benjamin Zhu, Kai Li, and R.~Hugo Patterson.
\newblock Avoiding the disk bottleneck in the data domain deduplication file
  system.
\newblock In Mary Baker and Erik Riedel, editors, \emph{6th {USENIX} Conference
  on File and Storage Technologies, {FAST} 2008, February 26-29, 2008, San
  Jose, CA, {USA}}, pages 269--282. {USENIX}, 2008.
\newblock URL \url{http://www.usenix.org/events/fast08/tech/zhu.html}.

\bibitem[{Wikipedia contributors}(2020)]{wiki:xxx}
{Wikipedia contributors}.
\newblock Twelvefold way --- {Wikipedia}{,} the free encyclopedia, 2020.
\newblock URL
  \url{https://en.wikipedia.org/w/index.php?title=Twelvefold_way&oldid=994636417}.
\newblock [Online; accessed 8-January-2021].

\bibitem[Peng et~al.(2017)Peng, Li, Song, and Liu]{DBLP:conf/aaai/PengLSL17}
Hao Peng, Jianxin Li, Yangqiu Song, and Yaopeng Liu.
\newblock Incrementally learning the hierarchical softmax function for neural
  language models.
\newblock In Satinder~P. Singh and Shaul Markovitch, editors, \emph{Proceedings
  of the Thirty-First {AAAI} Conference on Artificial Intelligence, February
  4-9, 2017, San Francisco, California, {USA}}, pages 3267--3273. {AAAI} Press,
  2017.
\newblock URL \url{http://aaai.org/ocs/index.php/AAAI/AAAI17/paper/view/14732}.

\bibitem[{Zhou} et~al.(2015){Zhou}, {Feng}, {Xia}, {Fu}, {Huang}, {Zhang}, and
  {Li}]{7208297}
Y.~{Zhou}, D.~{Feng}, W.~{Xia}, M.~{Fu}, F.~{Huang}, Y.~{Zhang}, and C.~{Li}.
\newblock Secdep: A user-aware efficient fine-grained secure deduplication
  scheme with multi-level key management.
\newblock In \emph{2015 31st Symposium on Mass Storage Systems and Technologies
  (MSST)}, pages 1--14, 2015.
\newblock \doi{10.1109/MSST.2015.7208297}.

\bibitem[Beller et~al.(2017)Beller, Gousios, and
  Zaidman]{DBLP:conf/msr/BellerGZ17a}
Moritz Beller, Georgios Gousios, and Andy Zaidman.
\newblock Travistorrent: synthesizing travis {CI} and github for full-stack
  research on continuous integration.
\newblock In Jes{\'{u}}s~M. Gonz{\'{a}}lez{-}Barahona, Abram Hindle, and Lin
  Tan, editors, \emph{Proceedings of the 14th International Conference on
  Mining Software Repositories, {MSR} 2017, Buenos Aires, Argentina, May 20-28,
  2017}, pages 447--450. {IEEE} Computer Society, 2017.
\newblock \doi{10.1109/MSR.2017.24}.
\newblock URL \url{https://doi.org/10.1109/MSR.2017.24}.

\end{thebibliography}






\end{document}